\documentclass[twocolumn,showpacs,aps,prb]{revtex4-1}
\usepackage{bm}
\usepackage{amssymb,amsmath}
\usepackage{graphicx}

\begin{document}

\title{Monte-Carlo simulations of the dissipative random transverse-field Ising chain}

\author{Manal Al-Ali}
\affiliation{Department of Physics, Missouri University of Science and Technology, Rolla, MO 65409, USA}

\author{Thomas Vojta}
\affiliation{Department of Physics, Missouri University of Science and Technology, Rolla, MO 65409, USA}

\begin{abstract}
We study the influence of Ohmic dissipation on the random transverse-field Ising chain by means of
large-scale Monte-Carlo simulations. To this end, we first map the Hamiltonian onto a classical
Ising model with long-range $1/\tau^2$ interaction in the time-like direction. We then apply the
highly efficient cluster algorithm proposed by Luijten and Bl\"ote for system with long-range interactions.
Our simulations show that Ohmic dissipation destroys the infinite-randomness quantum critical point of
the dissipationless system. Instead, the quantum phase transition between the paramagnetic and ferromagnetic
phases is smeared. We compare our results to recent predictions of a strong-disorder renormalization group
approach, and we discuss generalizations to higher dimensions as well as experiments.
\end{abstract}

\date{\today}

\maketitle

\section{Introduction}

Dissipation and disorder are two phenomena that can qualitatively change the properties
of quantum phase transitions. Dissipation alone can cause a finite-size quantum system
to undergo a transition. For example, the spin-boson model, a two-level system coupled
to a dissipative bath of harmonic oscillators, undergoes a quantum phase transition from
a fluctuating phase to a localized phase as the dissipation strength
increases.\cite{LCDFGZ87,Weiss_book93} Similar quantum phase transitions occur in other quantum
impurity models.\cite{VojtaM06}
In extended systems, the addition of dissipation can change the universality class of
the transition.\cite{WVTC05} Dissipation plays a particularly important role for quantum phase
transitions in metallic systems because the order parameter fluctuations are damped by the
coupling to gapless particle-hole excitations.\cite{Hertz76,Millis93,LRVW07}

Quenched disorder comprises impurities, defects, and other types of imperfections. It can
change the order of a transition from first-order to
continuous,\cite{ImryWortis79,HuiBerker89,AizenmanWehr89,GreenblattAizenmanLebowitz09}
and it can modify the critical behavior, resulting in  a different universality class.\cite{Harris74}
Moreover, at some quantum phase transitions, disorder leads to exotic exponential scaling\cite{Fisher92,Fisher95}
and to quantum Griffiths singularities \cite{ThillHuse95,RiegerYoung96} in the vicinity of the transition point
(see Refs.\ \onlinecite{Vojta06,Vojta10} for recent reviews).

If disorder and dissipation occur simultaneously in a system undergoing a quantum phase
transition, even stronger effects can be expected. The dissipative random transverse-field
Ising chain is a prototypical microscopic model for studying these phenomena. Due to the disorder,
this system contains rare large strongly coupled regions that are locally in the ferromagnetic phase
while the bulk system is still paramagnetic. Each of these locally ferromagnetic regions acts as
a quantum two-level system. In the presence of (Ohmic) dissipation,
the quantum dynamics of sufficiently large such regions completely freezes as they undergo
the localization transition of the Ohmic spin-boson model. Because each rare region freezes
independently from the rest of the system, the global quantum phase transition is
smeared.\cite{Vojta03a}

Going beyond these heuristic arguments, Schehr and Rieger\cite{SchehrRieger06,SchehrRieger08}
developed a numerical strong-disorder renormalization group approach to the dissipative random
transverse-field Ising chain. They confirmed the smeared transition scenario but focused on
the pseudo-critical point found at intermediate energies. Later, Hoyos and
Vojta\cite{HoyosVojta08,HoyosVojta12} developed a complete analytic theory by means of a slightly
modified renormalization group method. This theory becomes controlled in the strong-disorder
limit but its validity for weaker disorder requires independent verification.

In the present paper, we therefore perform large-scale Monte-Carlo simulations of the dissipative
random transverse-field Ising chain. Our goals are  to test the predictions of the strong-disorder
renormalization group theory of Refs.\ \onlinecite{HoyosVojta08,HoyosVojta12} and to determine to
what extent it applies to moderately or even weakly disordered systems. Our paper is organized as
follows. We define the quantum Hamiltonian in Sec.\ \ref{sec:model} and map it onto an anisotropic
two-dimensional classical Ising model. In Sec.\ \ref{sec:MC}, we describe our simulation method
and report the numerical results. We conclude in Sec.\ \ref{sec:conclusions} by discussing
generalizations to higher dimensions as well as experimental applications.

\section{Model and quantum-to-classical mapping}
\label{sec:model}

The Hamiltonian of the dissipative random transverse-field Ising chain consists of three parts,
\begin{equation}
H=H_{I}+H_{B}+H_{C}~.
\label{eq:H}
\end{equation}
$H_I$ denotes the Hamiltonian of the usual, dissipationless  transverse-field Ising model,
\begin{equation}
H_{I}=-\sum_{i}J_{i}\sigma_{i}^{z}\sigma_{i+1}^{z}-\sum_{i}h_{i}\sigma_{i}^{x}
\label{eq:H_I}
\end{equation}
where $\sigma_i^z$ and $\sigma_i^x$ are Pauli matrices representing the spin at lattice
site $i$. $J_i$ is the nearest-neighbor interaction between sites $i$ and $i+1$ while
$h_i$ is the transverse field acting on site $i$.

$H_B$ represents the Hamiltonians of independent harmonic oscillator baths (one for each site);
it is given by
\begin{equation}
H_{B}=\sum_{k,i}\omega_{k,i}\left(a_{k,i}^{\dagger}a_{k,i}^{\phantom{]\dagger}}+\frac{1}{2}\right)~.
\label{eq:H_B}
\end{equation}
Here, $\omega_{k,i}$ is the frequency of the $k$-th oscillator coupled to the spin
at site $i$, and $a_{k,i}^{\phantom{]\dagger}}$ and
$a_{k,i}^{\dagger}$ are the usual annihilation and creation operators.

The coupling between the spins and the dissipative baths is given by $H_C$ which reads
\begin{equation}
H_{C}=\sum_{i}\sigma_{i}^{z}\sum_{k}\lambda_{k,i}\left(a_{k,i}^{\dagger}+a_{k,i}^{\phantom{\dagger}}\right),
\label{eq:H-coupling}
\end{equation}
with $\lambda_{k,i}$ denoting the strength of the interaction.

The character and strength of the dissipation provided by the oscillator baths is
contained in their
spectral densities
\begin{equation}
{\cal E}_{i}(\omega)=\pi\sum_{k}\lambda_{k,i}^{2}\delta\left(\omega-\omega_{k,i}\right)~.
\label{eq:EE}
\end{equation}
Power-law spectral densities are of particular interest; they can be parameterized as
\begin{equation}
{\cal E}_{i}(\omega)= \frac{\pi}{2}\alpha_{i}\omega_{c}^{1-s}\omega^{s}\quad(\omega<\omega_{c})~.
\label{eq:spectral-function}
\end{equation}
Here, $\omega_{c}$ is a high-energy cutoff, and $\alpha_{i}$ is a dimensionless measure of the
dissipation strength. The value of the exponent $s$ determines the qualitative character of the dissipation.
Superohmic baths ($s>1$) are weak, they cannot induce a localization transition of a single spin.
The experimentally important Ohmic dissipation ($s=1$) constitutes the marginal case: If the dissipation
strength $\alpha$ is sufficiently large, an Ohmic baths can localize a single spin via a Kosterlitz-Thouless
impurity quantum phase transition. Subohmic dissipation
($s<1$) is even stronger, it also induces a single-spin localization transition. In this paper, we mostly
consider Ohmic dissipation, but we will comment on the other types in the concluding section. Moreover,
we restrict ourselves to the experimentally most interesting case of the bath cutoff $\omega_c$
being the largest energy, $\omega_c \gg h_i, J_i$.

As we are interested in the disordered, random version of the Hamiltonian (\ref{eq:H}), we allow the
interactions $J_i$, the transverse fields $h_i$, and the dissipation strengths $\alpha_i$ to be
independent random variables.

To apply our Monte-Carlo method, we now map the one-dimensional quantum Hamiltonian (\ref{eq:H})
onto a two-dimensional classical Ising model. This can be done using standard techniques, for example
using a Feynman path integral\cite{FeynmanHibbs_book65} representation of the partition function
or a transfer matrix method.\cite{Sachdev_book99} After integrating out all the bath oscillators,
we arrive at the following effective classical Hamiltonian:
\begin{eqnarray}
H_{cl} = &-&\sum_{i,\tau} J_i^x S_{i,\tau} S_{i+1,\tau} -\sum_{i,\tau} J_i^{\tau} S_{i,\tau} S_{i,\tau+1}\nonumber \\
         &-&\sum_{i,\tau,\tau'} \frac {\bar\alpha_i}{|\tau-\tau'|^{1+s}} S_{i,\tau} S_{i,\tau'}~.
\label{eq:H_cl}
\end{eqnarray}
Here, $S_{i,\tau}=\pm1$ are classical Ising variables, $i$ indexes the space direction and $\tau$ indexes
the imaginary time-like direction. The long-range interaction in the time direction in the last term results
from integrating out the dissipative baths. The coefficients $J_i^x$, $J_i^{\tau}$, and $\bar\alpha_i$ are
determined by the parameters of the original quantum Hamiltonian. In the following, we treat these coefficients
as fixed constants and drive the transition by varying the classical temperature $T$ (which is not identical
to the temperature of the original quantum system which is zero).

\section{Monte-Carlo simulations}
\label{sec:MC}
\subsection{Method and parameters}
\label{subsec:method}

We performed large-scale Monte Carlo simulations of the classical Hamiltonian (\ref{eq:H_cl})
for the case of Ohmic dissipation, $s=1$. To overcome the critical slowing down near the phase transition,
we used the Wolff cluster algorithm.\cite{Wolff89}

The long-range interaction in the time-like
direction (last term of the classical Hamiltonian (\ref{eq:H_cl})) poses additional problems.
A straightforward implementation of the Wolff algorithm for this Hamiltonian is not very efficient.
When building a cluster, all spins interacting with a given site need to be considered for addition to
the cluster, not just the nearest neighbor sites as in the case of short-range interactions. As
a result, the numerical effort scales quadratically with the number of sites in the time-like direction
rather then linearly. This problem is overcome by a clever version of the Wolff algorithm
due to Luijten and Bl\"ote\cite{LuijtenBlote95} that leads to linear scaling of the numerical effort with
system size, independent of the interaction range. We used this algorithm for all our simulations
(except for a few test runs in which we compared its results to that of straightforward implementations
of the Wolff and Metropolis algorithms).

We simulated systems with linear sizes of up $L=10000$ in space direction and $L_\tau= 6000$ in time direction. The results are
averages over large numbers of disorder realizations (from 200 to 2000 depending on system size). Each sample
was equilibrated using 200 Monte-Carlo sweeps (spin flips per site). After that, observables were measured
once every sweep for a total measurement period of 200 to 10000 sweeps, again depending on system size.

Quenched disorder was introduced into our simulations by making the interactions $J_i^x$ in the space direction
independent random variables governed by a binary probability distribution
\begin{equation}
P(J^x) = (1-p)\, \delta(J^{x} - 1) + p\, \delta(J^{x} - c)
\label{eq:binary}
\end{equation}
where $p$ is the concentration of weak bonds and $0<c\le 1$ is their interaction energy. We fixed these parameters at
$p=0.8$ and $c=0.25$. The interactions in time direction
were taken to be uniform $J_i^\tau\equiv J^\tau$, as were the dissipation strengths $\bar\alpha_i \equiv \bar\alpha$.

To test the predictions of the strong-disorder renormalization group theory,\cite{HoyosVojta08,HoyosVojta12}
we considered two different parameter sets. (i) Strong dissipation, $\bar\alpha=1$. In this case, we neglected the
short-range part of the interaction in the time direction (i.e., we set $J^\tau = 0$) as it is irrelevant for the critical
behavior. (ii) Weak dissipation. To study the crossover from the infinite-randomness criticality of the dissipationless
model, we set $J^\tau = 1$ and varied $\bar\alpha$ from 0 to 0.5. All simulations were performed on the Pegasus II computer
cluster at Missouri S\&T.

\subsection{Results for strong dissipation}
\label{subsec:results}

\begin{figure}
\includegraphics[width=8.5cm]{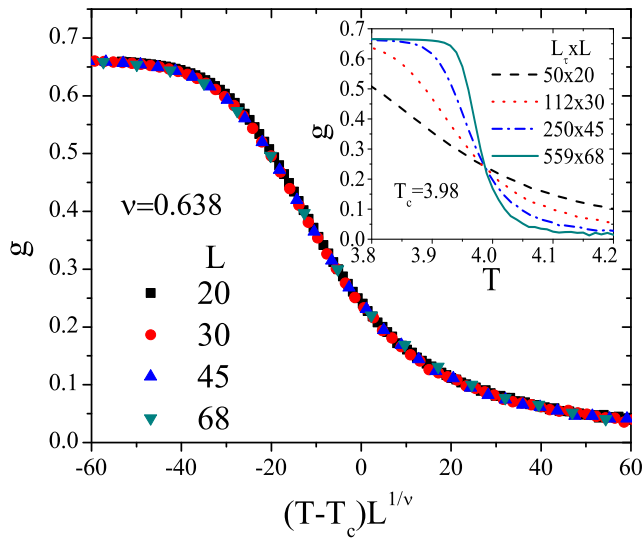}
\caption{(Color online) Finite-size scaling of the Binder cumulant $g$ for the classical Hamiltonian
       (\ref{eq:H_cl}) with $\bar\alpha=1$ and $J^\tau=0$ in the clean limit $p=0$ giving a correlation length critical exponent $\nu=0.638$. The inset
       shows the raw data which give a high-quality crossing at $T_c^0\approx 3.98$. The sample shapes
       ($L$ vs. $L_\tau$) reflect the dynamical exponent value $z=1.98$.}
\label{fig:cleanbinder}
\end{figure}

\begin{figure}
\includegraphics[width=8.5cm]{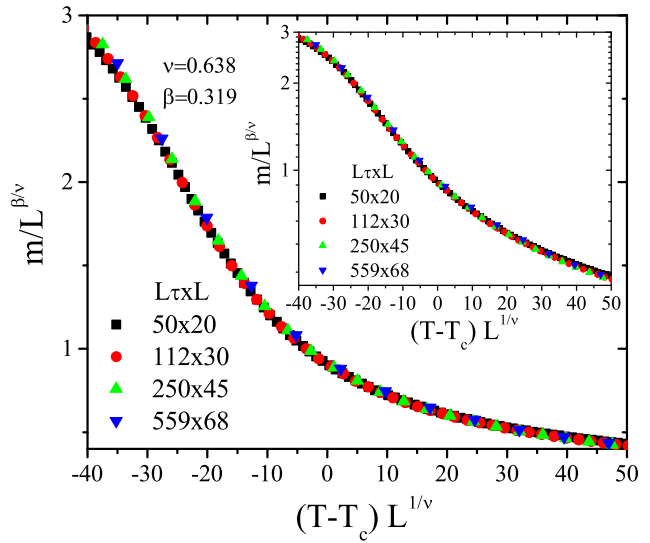}
\caption{(Color online) Finite-size scaling of the magnetization $m$ for the classical Hamiltonian
       (\ref{eq:H_cl}) with $\bar\alpha=1$ and $J^\tau=0$ in the clean limit $p=0$ giving an order parameter critical exponent $\beta=0.319$. The inset
       shows the same data on a logarithmic scale.}
\label{fig:cleanmag}
\end{figure}

In this section we discuss results for the case $\bar\alpha=1$ and $J^\tau=0$. To test our implementation of the
Luijten-Bl\"ote algorithm,\cite{LuijtenBlote95} we first considered a clean system with zero concentration of
weak bonds ($p=0$). We analyzed the finite-size scaling behavior of the magnetization $m$, the magnetic susceptibility
$\chi$ as well as the Binder cumulant $g=1-\langle m^4 \rangle/(3\langle m^2 \rangle^2)$ close to the transition
temperature $T_c^0\approx 3.98$. Results for the Binder
cumulant and the magnetization are presented in Figs.\ \ref{fig:cleanbinder} and \ref{fig:cleanmag}.
Both quantities display high-quality scaling as does the susceptibility (not shown).  The resulting critical exponents,
$\nu=0.638$, $z=1.98$, $\beta=0.319$, and $\gamma=1.27$ agree with literature values for the dissipative transverse-field
Ising chain.\cite{WVTC05}

We note that the correlation length exponent violates the Harris criterion\cite{Harris74}
$d_\perp\nu > 2$. Here, $d_\perp=1$ is the number of ``random dimensions'' which differs from the total dimensionality
$d=2$ of the classical model (\ref{eq:H_cl}) because the disorder is perfectly correlated in the time-like
direction. The violation of Harris' inequality suggests that weak disorder is a relevant perturbation at the clean critical point;
the character of the transition is thus expected to change upon the introduction of disorder.

In addition to providing a test of our numerical algorithm, the clean system simulations also give us a value for
the upper Griffiths temperature $T_u$ for later use in the analysis of the disordered case. The upper Griffiths
temperature is the temperature above which no (rare) locally ordered regions can exist in the disordered system.
For the binary disorder distribution (\ref{eq:binary}), the upper Griffiths temperature is identical to the
critical temperature of an impurity-free system ($p=0$). Thus, in our case $T_u=T_c^0\approx 3.98$.

We now turn to our simulations of the disordered case, using $p=0.8$ and $c=0.25$ in the binary distribution
(\ref{eq:binary}). To establish the smeared character of the phase transition, we analyzed the temperature
dependence of the magnetization. According to the theoretical predictions,\cite{Vojta03a,Vojta03b} the magnetization
is expected to develop an exponential tail of the form
\begin{equation}
m = m_0 \, \exp[-(T_c^0-T)^{-\nu}]
\label{eq:m-tail}
\end{equation}
towards the upper Griffiths temperature $T_u=T_c^0$. Here, $\nu$ is the correlation length exponent of the clean system.
This tail forms because sufficiently large individual rare regions
undergo the phase transition independently at different values of the tuning parameter.
(After the quantum-to-classical mapping, these rare regions correspond to
``strips'' of finite width in the space direction.)
To see this phenomenon in the simulations of finite-size systems requires a careful choice
of the simulation parameters. In particular, the system size $L_\tau$ in the time-like direction needs to be very large
to allow for sharp transitions of the individual rare regions to occur. Note that the smeared transition
in the original quantum Hamiltonian (\ref{eq:H}) occurs only in the zero-temperature limit which corresponds to the
limit $L_\tau \to \infty$ in the classical model (\ref{eq:H_cl}). In contrast, the system size $L$ is space direction
is not very important because the tail of the smeared transition is produced by finite-size rare regions (and the spatial correlation length
remains finite).

\begin{figure}
\includegraphics[width=8.5cm]{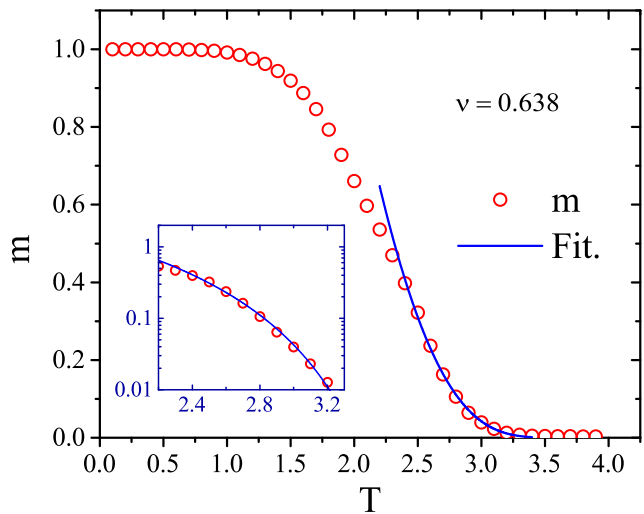}
\caption{(Color online) Magnetization $m$ vs temperature $T$ for the classical Hamiltonian
       (\ref{eq:H_cl}) with $p=0.8$, $c=0.25$, $\bar\alpha=1$ and $J^\tau=0$ for a system of size $L=50, L_\tau=6000$,
       averaged over 200 disorder realizations. $m$ develops a pronounced tail towards $T_c^0=3.98$.
       The solid line is a fit to (\ref{eq:m-tail}).  The semi-log plot of the same data
       in the inset shows that the theoretical
       prediction fits the tail region for almost two orders of magnitude in $m$.}
\label{fig:mvsTstrong}
\end{figure}

Figure \ref{fig:mvsTstrong} shows the magnetization as a function of temperature for a system of size $L=50, L_\tau=6000$,
averaged over 200 disorder realizations.
The data display a pronounced tail towards the upper Griffiths temperature $T_u=T_c^0$. We have compared
different system sizes to ensure that this tail is \emph{not} the result of any remaining finite-size effects.
To compare with the theoretical predictions, we fit the magnetization in the tail region (temperatures
above the inflection point at $T\approx 2.3$) to the exponential form (\ref{eq:m-tail}). The numerical data
follow the prediction for almost two orders of magnitude in $m$ (temperatures between 2.3 and 3.2). At
higher temperatures, the numerical magnetization value is dominated by Monte-Carlo noise and thus
saturates at a roughly temperature-independent value. (To suppress this effect, one would need to use
even larger system sizes.)

In addition to the magnetization, we also studied the magnetic susceptibility in the tail region
of the smeared transition. According to the strong-disorder renormalization group theory,\cite{HoyosVojta08,HoyosVojta12}
the temperature dependence of the susceptibility of the quantum Hamiltonian (\ref{eq:H}) is
characterized by a complicated double crossover (see Fig.\ 3b of Ref.\ \onlinecite{HoyosVojta12}).
At higher temperatures, the physics is dominated by
small clusters that cannot order (or freeze) independently. Thus, they display
power-law quantum Griffiths behavior similar to the dissipationless
system. At lower temperatures, the relevant clusters become large enough to undergo
the localization phase transition independently, i.e., their quantum dynamics freezes.
As a result, each such region makes a classical Curie contribution to the susceptibility.

\begin{figure}
\includegraphics[width=8.5cm]{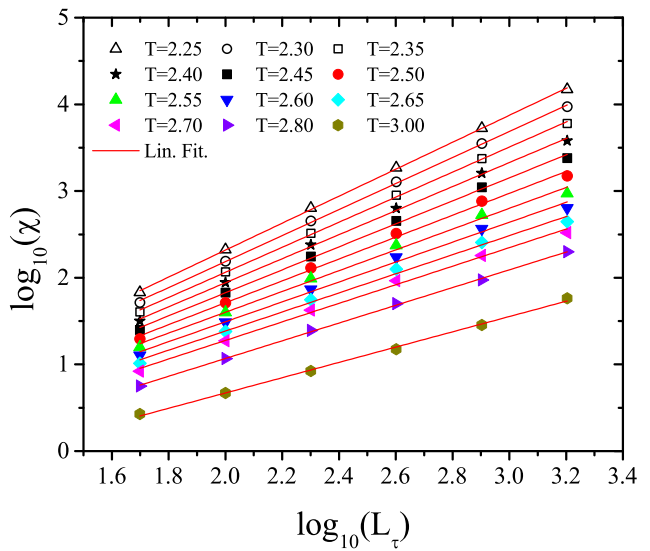}
\caption{(Color online) Susceptibility $\chi$ vs system size $L_\tau$ for the classical Hamiltonian
       (\ref{eq:H_cl}) with $p=0.8$, $c=0.25$, $J^\tau=0$, and $\bar\alpha=1$  at different values of the classical temperature $T$.
       The spatial system size is $L=3000$. The solid lines are fits to the power-law (\ref{eq:chivsLt}).}
\label{fig:chivsLtstrong}
\end{figure}

Under the quantum-to-classical mapping, the (inverse) temperature in the
quantum Hamiltonian (\ref{eq:H}) maps onto the time-like system size $L_\tau$ in the classical model
(\ref{eq:H_cl}). Figure \ref{fig:chivsLtstrong} thus shows the dependence of the
magnetic susceptibility $\chi$  on $L_\tau$ for several values of the classical temperature $T$
in the tail region of the smeared transition.
The data can all be fitted well by the power-law relation
\begin{equation}
\chi \sim L_\tau^{1-\lambda}=L_\tau^{1\pm 1/z'}
\label{eq:chivsLt}
\end{equation}
where $\lambda$ is the usual nonuniversal Griffiths exponent (see, e.g., Ref.\ \onlinecite{Vojta06})
and $z'$ is the corresponding dynamical exponent in the Griffiths phase. Here, the $+$ sign in the exponent
applies in the ferromagnetic Griffiths phase and the $-$ sign in the paramagnetic Griffiths phase.
For the fit curves in Fig.\ \ref{fig:chivsLtstrong}, $\lambda$ ranges from -0.55 at $T=2.25$ to
0.12 at $T=3.0$.

The fact that all data in Fig.\ \ref{fig:chivsLtstrong} follow (pure) power laws with a monotonously
changing exponent $\lambda$ suggests that our simulations are still in the transient Griffiths regime
predicted by the strong-disorder renormalization group. They have not yet reached the asymptotic
large-$L_\tau$ regime dominated by frozen clusters. In fact, the data at the highest classical temperature
$T=3.0$ show a slight upturn for large $L_\tau$ which may indicate the beginning of the crossover to the
asymptotic regime.

\subsection{Crossover between the dissipationless and dissipative cases}
\label{subsec:crossover}

\begin{figure}
\includegraphics[width=8.5cm]{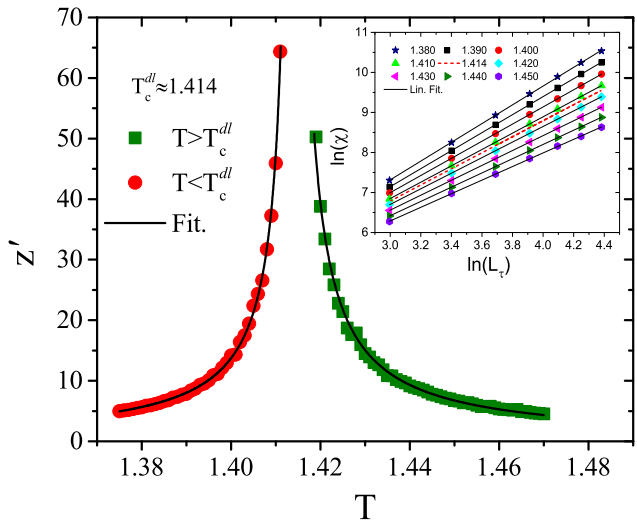}
\caption{(Color online) Griffiths dynamical exponent $z'$ vs temperature $T$
       for the classical Hamiltonian (\ref{eq:H_cl}) with $p=0.8$, $c=0.25$, and $J^\tau=1$ in the absence
       of dissipation ($\bar\alpha=0)$. A fit to the expected\cite{Fisher95} power law $z' \sim |T-T_c^{dl}|^{-1}$
       results in $T_c^{dl}\approx 1.414$. The inset shows the raw susceptibility data as a function of the
       time-like system size $L_\tau$. The spatial system size is $L=2000$, and the data are averaged
       over 1400 to 2000 disorder realizations. }
\label{fig:zvsTdissipationless}
\end{figure}

The strong-disorder renormalization group theory\cite{HoyosVojta08,HoyosVojta12} also makes detailed
predictions for the crossover from the dissipationless to the dissipative behavior with increasing
dissipation strength $\alpha$. To investigate this crossover numerically, we first analyzed a dissipationless
system by setting $\bar\alpha=0$ and $J^\tau=1$. In this case, the theory predicts a sharp transition
governed by an infinite-randomness critical point.\cite{Fisher92,Fisher95} We confirmed this prediction by
applying the methods of Ref.\ \onlinecite{HrahshehBarghathiVojta11} to the case at hand, in agreement with
earlier simulation results in the literature.\cite{YoungRieger96}  Specifically, by analyzing
the finite-size scaling properties of the susceptibility, we found the critical temperature
of the dissipationless system to be $T_c^{dl}\approx 1.414$ (see Fig.\ \ref{fig:zvsTdissipationless}).

\begin{figure}
\includegraphics[width=8.5cm]{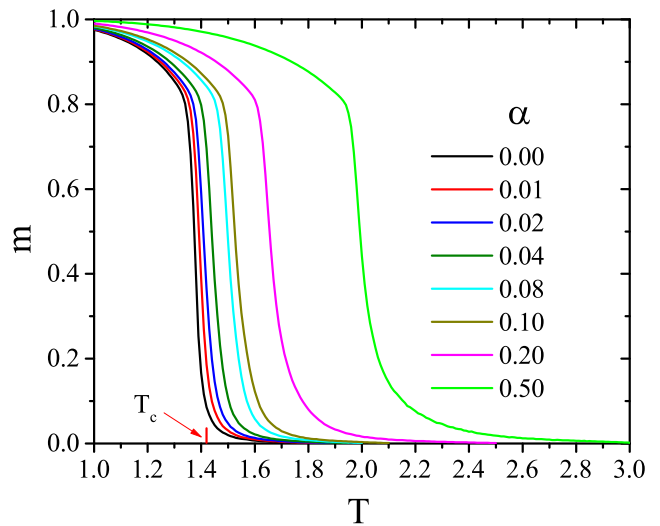}
\caption{(Color online) Magnetization $m$ vs temperature $T$ for the classical Hamiltonian
       (\ref{eq:H_cl}) with $p=0.8$, $c=0.25$, and $J^\tau=1$ for several values of the
       dissipation strength $\bar\alpha$. The system size is $L=200, L_\tau=10000$, and the data are averaged
       over 500 disorder realizations. The critical temperature of the dissipationless
       system ($\bar\alpha=0$) is $T_c^{dl}\approx 1.414$.}
\label{fig:mvsTweak}
\end{figure}

We then performed simulations for $J^\tau=1$ and several values of the dissipation strength between
$\bar\alpha= 0.01$ and 0.5. The resulting magnetization in the temperature range $T=1.0$ to 3.0 is
presented in Fig.\ \ref{fig:mvsTweak}.
In this figure, even the magnetization of the dissipationless system ($\bar\alpha=0$), which has
a sharp phase transition in the thermodynamic limit, shows a small ``tail.'' It stems from the remaining finite-size effects
and can thus not be completely avoided. With increasing dissipation, the magnetization tail becomes
much more pronounced than this finite-size tail, again lending support to the smeared transition scenario of Refs.\
\onlinecite{HoyosVojta08,HoyosVojta12}.

However, a quantitative comparison with the theory of
the crossover between the dissipationless and dissipative cases would require analyzing the
weak-dissipation data ($\bar\alpha\ll 1$). For these cases, the smearing-induced magnetization
tail is masked by the remaining finite-size effects and can thus not be studied quantitatively.
Analogous problems also hinder the analysis of the magnetic susceptibility. We conclude that
although our weak-dissipation results are in qualitative agreement with the theoretical predictions,
a quantitative test of the crossover would require significantly larger systems.

\section{Conclusions}
\label{sec:conclusions}

To summarize, we investigated the quantum phase transition of a random transverse-field
Ising chain in the presence of Ohmic dissipation. To this end, we first mapped the quantum Hamiltonian onto a
classical two-dimensional Ising model with long-range ($1/\tau^2$) interactions in the time-like direction.
This classical system was then studied by means of Monte-Carlo simulations using the
Luijten/Bl\"ote version of the Wolff cluster algorithm that efficiently deals with the long-range interactions.

Our results provide numerical evidence for the predictions of a recent strong-disorder renormalization
group theory\cite{HoyosVojta08,HoyosVojta12} as well as earlier heuristic arguments.\cite{Vojta03a}
In particular, the simulations confirm that the combined effects
of disorder and dissipation lead to a destruction of the sharp quantum phase transition by smearing.
This happens because  different spatial regions can undergo the phase transition independently
of the bulk system at different values of the tuning parameter.

For sufficiently strong dissipation (here, $\bar\alpha=1$), we could quantitatively compare the
simulation data with the theoretical predictions and found them in good agreement. For weak dissipation,
a quantitative comparison was not possible because the dissipation-induced tail of the smeared transition
is small and thus masked by the remaining finite-size effects in our simulations.

As pointed out in the introduction, the renormalization group theory\cite{HoyosVojta08,HoyosVojta12}
becomes controlled in the limit of strong randomness while its applicability to weak and moderate disorder
requires independent verification. The binary distribution (\ref{eq:binary}) used in our simulations
constitutes moderate disorder, because $\Delta J^x / J^x$ is of order unity but the distribution is \emph{not}
broad on a logarithmic scale. Our simulations thus show that a moderately disordered system follows
the predictions of the strong-disorder theory. Moreover, because the clean system violates the Harris
criterion (see Sec.\ \ref{subsec:results}) weak (bare) disorder will increase under coarse graining.
This strongly suggests that the strong-disorder renormalization group theory governs the transition
for any nonzero disorder strength. A direct numerical verification for weak disorder would be computationally
expensive because the crossover to the disorder-dominated behavior would occur at very large
system sizes only.

Both the renormalization group theory and the present simulations address the case of one space dimension.
However, many applications of the smeared-transition scenario are actually in higher-dimensional systems.
It is thus useful to discuss what changes in higher dimensions. The most important insight is that the
smearing of the transition is driven by the freezing of individual \emph{finite-size} regions of the sample.
This implies that the space dimensionality does not play an important role. We thus expect that the
same smeared-transition scenario applies in all dimensions. To test this numerically, one could map the
$d$-dimensional dissipative random transverse-field Ising model to a $(d+1)$-dimensional version of the
classical Hamiltonian (\ref{eq:H_cl}) and then apply the methods of this paper. Generalizations to other
types of dissipation (subohmic and superohmic) are also straight forward, they simply lead to different 
power-laws in the long-range interaction in the classical Hamiltonian (\ref{eq:H_cl}). The  
Luijten-Bl\"ote algorithm\cite{LuijtenBlote95} can be applied in all of these cases.

The most important experimental realizations of smeared quantum phase transitions can arguably be found
in disordered metallic magnets. The standard approach to magnetic quantum phase transitions in Fermi
liquids\cite{Hertz76,Millis93} leads to an order-parameter field theory with a structure similar to
our classical Hamiltonian (\ref{eq:H_cl}). In particular, the order-parameter fluctuations experience
Ohmic damping reflected in a long-range $1/\tau^2$ interaction in the imaginary time direction. Recently,
indications of frozen local clusters have been observed\cite{UbaidKassisVojtaSchroeder10,SchroederUbaidKassisVojta11}
near the ferromagnetic quantum phase transition in ${\mathrm{Ni}}_{1-x}{\mathrm{V}}_x$. Moreover, the ferromagnetic quantum
phase transition in ${\mathrm{Sr}}_{1-x}{\mathrm{Ca}}_{x}{\mathrm{RuO}}_{3}$ was shown to be smeared
by the disorder introduced via the Ca substitution.\cite{Demkoetal12}

\section*{Acknowledgements}

We acknowledge useful discussions with Jos\'e Hoyos. 
This work has been supported in part by the NSF under grant nos.\ DMR-1205803
and PHYS-1066293 as well as the hospitality of the Aspen Center for Physics.

\bibliographystyle{apsrev4-1}
\bibliography{../00Bibtex/rareregions}

\begin{thebibliography}{33}%
\makeatletter
\providecommand \@ifxundefined [1]{%
 \@ifx{#1\undefined}
}%
\providecommand \@ifnum [1]{%
 \ifnum #1\expandafter \@firstoftwo
 \else \expandafter \@secondoftwo
 \fi
}%
\providecommand \@ifx [1]{%
 \ifx #1\expandafter \@firstoftwo
 \else \expandafter \@secondoftwo
 \fi
}%
\providecommand \natexlab [1]{#1}%
\providecommand \enquote  [1]{``#1''}%
\providecommand \bibnamefont  [1]{#1}%
\providecommand \bibfnamefont [1]{#1}%
\providecommand \citenamefont [1]{#1}%
\providecommand \href@noop [0]{\@secondoftwo}%
\providecommand \href [0]{\begingroup \@sanitize@url \@href}%
\providecommand \@href[1]{\@@startlink{#1}\@@href}%
\providecommand \@@href[1]{\endgroup#1\@@endlink}%
\providecommand \@sanitize@url [0]{\catcode `\\12\catcode `\$12\catcode
  `\&12\catcode `\#12\catcode `\^12\catcode `\_12\catcode `\%12\relax}%
\providecommand \@@startlink[1]{}%
\providecommand \@@endlink[0]{}%
\providecommand \url  [0]{\begingroup\@sanitize@url \@url }%
\providecommand \@url [1]{\endgroup\@href {#1}{\urlprefix }}%
\providecommand \urlprefix  [0]{URL }%
\providecommand \Eprint [0]{\href }%
\providecommand \doibase [0]{http://dx.doi.org/}%
\providecommand \selectlanguage [0]{\@gobble}%
\providecommand \bibinfo  [0]{\@secondoftwo}%
\providecommand \bibfield  [0]{\@secondoftwo}%
\providecommand \translation [1]{[#1]}%
\providecommand \BibitemOpen [0]{}%
\providecommand \bibitemStop [0]{}%
\providecommand \bibitemNoStop [0]{.\EOS\space}%
\providecommand \EOS [0]{\spacefactor3000\relax}%
\providecommand \BibitemShut  [1]{\csname bibitem#1\endcsname}%
\let\auto@bib@innerbib\@empty
\bibitem [{\citenamefont {Leggett}\ \emph {et~al.}(1987)\citenamefont
  {Leggett}, \citenamefont {Chakravarty}, \citenamefont {Dorsey}, \citenamefont
  {Fisher}, \citenamefont {Garg},\ and\ \citenamefont {Zwerger}}]{LCDFGZ87}%
  \BibitemOpen
  \bibfield  {author} {\bibinfo {author} {\bibfnamefont {A.~J.}\ \bibnamefont
  {Leggett}}, \bibinfo {author} {\bibfnamefont {S.}~\bibnamefont
  {Chakravarty}}, \bibinfo {author} {\bibfnamefont {A.~T.}\ \bibnamefont
  {Dorsey}}, \bibinfo {author} {\bibfnamefont {M.~P.~A.}\ \bibnamefont
  {Fisher}}, \bibinfo {author} {\bibfnamefont {A.}~\bibnamefont {Garg}}, \ and\
  \bibinfo {author} {\bibfnamefont {W.}~\bibnamefont {Zwerger}},\ }\href@noop
  {} {\bibfield  {journal} {\bibinfo  {journal} {Rev. Mod. Phys.}\ }\textbf
  {\bibinfo {volume} {59}},\ \bibinfo {pages} {1} (\bibinfo {year}
  {1987})}\BibitemShut {NoStop}%
\bibitem [{\citenamefont {Weiss}(1993)}]{Weiss_book93}%
  \BibitemOpen
  \bibfield  {author} {\bibinfo {author} {\bibfnamefont {U.}~\bibnamefont
  {Weiss}},\ }\href@noop {} {\emph {\bibinfo {title} {Quantum disspative
  systems}}}\ (\bibinfo  {publisher} {World Scientific},\ \bibinfo {address}
  {Singapore},\ \bibinfo {year} {1993})\BibitemShut {NoStop}%
\bibitem [{\citenamefont {Vojta}(2006{\natexlab{a}})}]{VojtaM06}%
  \BibitemOpen
  \bibfield  {author} {\bibinfo {author} {\bibfnamefont {M.}~\bibnamefont
  {Vojta}},\ }\href@noop {} {\bibfield  {journal} {\bibinfo  {journal} {Phil.
  Mag.}\ }\textbf {\bibinfo {volume} {86}},\ \bibinfo {pages} {1807} (\bibinfo
  {year} {2006}{\natexlab{a}})}\BibitemShut {NoStop}%
\bibitem [{\citenamefont {Werner}\ \emph {et~al.}(2005)\citenamefont {Werner},
  \citenamefont {V{\"o}lker}, \citenamefont {Troyer},\ and\ \citenamefont
  {Chakravarty}}]{WVTC05}%
  \BibitemOpen
  \bibfield  {author} {\bibinfo {author} {\bibfnamefont {P.}~\bibnamefont
  {Werner}}, \bibinfo {author} {\bibfnamefont {K.}~\bibnamefont {V{\"o}lker}},
  \bibinfo {author} {\bibfnamefont {M.}~\bibnamefont {Troyer}}, \ and\ \bibinfo
  {author} {\bibfnamefont {S.}~\bibnamefont {Chakravarty}},\ }\href@noop {}
  {\bibfield  {journal} {\bibinfo  {journal} {Phys. Rev. Lett.}\ }\textbf
  {\bibinfo {volume} {94}},\ \bibinfo {pages} {047201} (\bibinfo {year}
  {2005})}\BibitemShut {NoStop}%
\bibitem [{\citenamefont {Hertz}(1976)}]{Hertz76}%
  \BibitemOpen
  \bibfield  {author} {\bibinfo {author} {\bibfnamefont {J.}~\bibnamefont
  {Hertz}},\ }\href@noop {} {\bibfield  {journal} {\bibinfo  {journal} {Phys.
  Rev. B}\ }\textbf {\bibinfo {volume} {14}},\ \bibinfo {pages} {1165}
  (\bibinfo {year} {1976})}\BibitemShut {NoStop}%
\bibitem [{\citenamefont {Millis}(1993)}]{Millis93}%
  \BibitemOpen
  \bibfield  {author} {\bibinfo {author} {\bibfnamefont {A.~J.}\ \bibnamefont
  {Millis}},\ }\href@noop {} {\bibfield  {journal} {\bibinfo  {journal} {Phys.
  Rev. B}\ }\textbf {\bibinfo {volume} {48}},\ \bibinfo {pages} {7183}
  (\bibinfo {year} {1993})}\BibitemShut {NoStop}%
\bibitem [{\citenamefont {von L{\"o}hneysen}\ \emph {et~al.}(2007)\citenamefont
  {von L{\"o}hneysen}, \citenamefont {Rosch}, \citenamefont {Vojta},\ and\
  \citenamefont {W{\"o}lfle}}]{LRVW07}%
  \BibitemOpen
  \bibfield  {author} {\bibinfo {author} {\bibfnamefont {H.}~\bibnamefont {von
  L{\"o}hneysen}}, \bibinfo {author} {\bibfnamefont {A.}~\bibnamefont {Rosch}},
  \bibinfo {author} {\bibfnamefont {M.}~\bibnamefont {Vojta}}, \ and\ \bibinfo
  {author} {\bibfnamefont {P.}~\bibnamefont {W{\"o}lfle}},\ }\href@noop {}
  {\bibfield  {journal} {\bibinfo  {journal} {Rev. Mod. Phys.}\ }\textbf
  {\bibinfo {volume} {79}},\ \bibinfo {pages} {1015} (\bibinfo {year}
  {2007})}\BibitemShut {NoStop}%
\bibitem [{\citenamefont {Imry}\ and\ \citenamefont
  {Wortis}(1979)}]{ImryWortis79}%
  \BibitemOpen
  \bibfield  {author} {\bibinfo {author} {\bibfnamefont {Y.}~\bibnamefont
  {Imry}}\ and\ \bibinfo {author} {\bibfnamefont {M.}~\bibnamefont {Wortis}},\
  }\href@noop {} {\bibfield  {journal} {\bibinfo  {journal} {Phys. Rev. B}\
  }\textbf {\bibinfo {volume} {19}},\ \bibinfo {pages} {3580} (\bibinfo {year}
  {1979})}\BibitemShut {NoStop}%
\bibitem [{\citenamefont {Hui}\ and\ \citenamefont
  {Berker}(1989)}]{HuiBerker89}%
  \BibitemOpen
  \bibfield  {author} {\bibinfo {author} {\bibfnamefont {K.}~\bibnamefont
  {Hui}}\ and\ \bibinfo {author} {\bibfnamefont {A.~N.}\ \bibnamefont
  {Berker}},\ }\href@noop {} {\bibfield  {journal} {\bibinfo  {journal} {Phys.
  Rev. Lett.}\ }\textbf {\bibinfo {volume} {62}},\ \bibinfo {pages} {2507}
  (\bibinfo {year} {1989})}\BibitemShut {NoStop}%
\bibitem [{\citenamefont {Aizenman}\ and\ \citenamefont
  {Wehr}(1989)}]{AizenmanWehr89}%
  \BibitemOpen
  \bibfield  {author} {\bibinfo {author} {\bibfnamefont {M.}~\bibnamefont
  {Aizenman}}\ and\ \bibinfo {author} {\bibfnamefont {J.}~\bibnamefont
  {Wehr}},\ }\href@noop {} {\bibfield  {journal} {\bibinfo  {journal} {Phys.
  Rev. Lett.}\ }\textbf {\bibinfo {volume} {62}},\ \bibinfo {pages} {2503}
  (\bibinfo {year} {1989})}\BibitemShut {NoStop}%
\bibitem [{\citenamefont {Greenblatt}\ \emph {et~al.}(2009)\citenamefont
  {Greenblatt}, \citenamefont {Aizenman},\ and\ \citenamefont
  {Lebowitz}}]{GreenblattAizenmanLebowitz09}%
  \BibitemOpen
  \bibfield  {author} {\bibinfo {author} {\bibfnamefont {R.~L.}\ \bibnamefont
  {Greenblatt}}, \bibinfo {author} {\bibfnamefont {M.}~\bibnamefont
  {Aizenman}}, \ and\ \bibinfo {author} {\bibfnamefont {J.~L.}\ \bibnamefont
  {Lebowitz}},\ }\href {\doibase 10.1103/PhysRevLett.103.197201} {\bibfield
  {journal} {\bibinfo  {journal} {Phys. Rev. Lett.}\ }\textbf {\bibinfo
  {volume} {103}},\ \bibinfo {pages} {197201} (\bibinfo {year}
  {2009})}\BibitemShut {NoStop}%
\bibitem [{\citenamefont {Harris}(1974)}]{Harris74}%
  \BibitemOpen
  \bibfield  {author} {\bibinfo {author} {\bibfnamefont {A.~B.}\ \bibnamefont
  {Harris}},\ }\href@noop {} {\bibfield  {journal} {\bibinfo  {journal} {J.
  Phys. C}\ }\textbf {\bibinfo {volume} {7}},\ \bibinfo {pages} {1671}
  (\bibinfo {year} {1974})}\BibitemShut {NoStop}%
\bibitem [{\citenamefont {Fisher}(1992)}]{Fisher92}%
  \BibitemOpen
  \bibfield  {author} {\bibinfo {author} {\bibfnamefont {D.~S.}\ \bibnamefont
  {Fisher}},\ }\href@noop {} {\bibfield  {journal} {\bibinfo  {journal} {Phys.
  Rev. Lett.}\ }\textbf {\bibinfo {volume} {69}},\ \bibinfo {pages} {534}
  (\bibinfo {year} {1992})}\BibitemShut {NoStop}%
\bibitem [{\citenamefont {Fisher}(1995)}]{Fisher95}%
  \BibitemOpen
  \bibfield  {author} {\bibinfo {author} {\bibfnamefont {D.~S.}\ \bibnamefont
  {Fisher}},\ }\href@noop {} {\bibfield  {journal} {\bibinfo  {journal} {Phys.
  Rev. B}\ }\textbf {\bibinfo {volume} {51}},\ \bibinfo {pages} {6411}
  (\bibinfo {year} {1995})}\BibitemShut {NoStop}%
\bibitem [{\citenamefont {Thill}\ and\ \citenamefont
  {Huse}(1995)}]{ThillHuse95}%
  \BibitemOpen
  \bibfield  {author} {\bibinfo {author} {\bibfnamefont {M.}~\bibnamefont
  {Thill}}\ and\ \bibinfo {author} {\bibfnamefont {D.~A.}\ \bibnamefont
  {Huse}},\ }\href@noop {} {\bibfield  {journal} {\bibinfo  {journal} {Physica
  A}\ }\textbf {\bibinfo {volume} {214}},\ \bibinfo {pages} {321} (\bibinfo
  {year} {1995})}\BibitemShut {NoStop}%
\bibitem [{\citenamefont {Rieger}\ and\ \citenamefont
  {Young}(1996)}]{RiegerYoung96}%
  \BibitemOpen
  \bibfield  {author} {\bibinfo {author} {\bibfnamefont {H.}~\bibnamefont
  {Rieger}}\ and\ \bibinfo {author} {\bibfnamefont {A.~P.}\ \bibnamefont
  {Young}},\ }\href@noop {} {\bibfield  {journal} {\bibinfo  {journal} {Phys.
  Rev. B}\ }\textbf {\bibinfo {volume} {54}},\ \bibinfo {pages} {3328}
  (\bibinfo {year} {1996})}\BibitemShut {NoStop}%
\bibitem [{\citenamefont {Vojta}(2006{\natexlab{b}})}]{Vojta06}%
  \BibitemOpen
  \bibfield  {author} {\bibinfo {author} {\bibfnamefont {T.}~\bibnamefont
  {Vojta}},\ }\href@noop {} {\bibfield  {journal} {\bibinfo  {journal} {J.
  Phys. A}\ }\textbf {\bibinfo {volume} {39}},\ \bibinfo {pages} {R143}
  (\bibinfo {year} {2006}{\natexlab{b}})}\BibitemShut {NoStop}%
\bibitem [{\citenamefont {Vojta}(2010)}]{Vojta10}%
  \BibitemOpen
  \bibfield  {author} {\bibinfo {author} {\bibfnamefont {T.}~\bibnamefont
  {Vojta}},\ }\href@noop {} {\bibfield  {journal} {\bibinfo  {journal} {J. Low
  Temp. Phys.}\ }\textbf {\bibinfo {volume} {161}},\ \bibinfo {pages} {299}
  (\bibinfo {year} {2010})}\BibitemShut {NoStop}%
\bibitem [{\citenamefont {Vojta}(2003{\natexlab{a}})}]{Vojta03a}%
  \BibitemOpen
  \bibfield  {author} {\bibinfo {author} {\bibfnamefont {T.}~\bibnamefont
  {Vojta}},\ }\href@noop {} {\bibfield  {journal} {\bibinfo  {journal} {Phys.
  Rev. Lett.}\ }\textbf {\bibinfo {volume} {90}},\ \bibinfo {pages} {107202}
  (\bibinfo {year} {2003}{\natexlab{a}})}\BibitemShut {NoStop}%
\bibitem [{\citenamefont {Schehr}\ and\ \citenamefont
  {Rieger}(2006)}]{SchehrRieger06}%
  \BibitemOpen
  \bibfield  {author} {\bibinfo {author} {\bibfnamefont {G.}~\bibnamefont
  {Schehr}}\ and\ \bibinfo {author} {\bibfnamefont {H.}~\bibnamefont
  {Rieger}},\ }\href@noop {} {\bibfield  {journal} {\bibinfo  {journal} {Phys.
  Rev. Lett.}\ }\textbf {\bibinfo {volume} {96}},\ \bibinfo {pages} {227201}
  (\bibinfo {year} {2006})}\BibitemShut {NoStop}%
\bibitem [{\citenamefont {Schehr}\ and\ \citenamefont
  {Rieger}(2008)}]{SchehrRieger08}%
  \BibitemOpen
  \bibfield  {author} {\bibinfo {author} {\bibfnamefont {G.}~\bibnamefont
  {Schehr}}\ and\ \bibinfo {author} {\bibfnamefont {H.}~\bibnamefont
  {Rieger}},\ }\href@noop {} {\bibfield  {journal} {\bibinfo  {journal} {J.
  Stat. Mech.}\ }\textbf {\bibinfo {volume} {2008}},\ \bibinfo {pages} {P04012}
  (\bibinfo {year} {2008})}\BibitemShut {NoStop}%
\bibitem [{\citenamefont {Hoyos}\ and\ \citenamefont
  {Vojta}(2008)}]{HoyosVojta08}%
  \BibitemOpen
  \bibfield  {author} {\bibinfo {author} {\bibfnamefont {J.~A.}\ \bibnamefont
  {Hoyos}}\ and\ \bibinfo {author} {\bibfnamefont {T.}~\bibnamefont {Vojta}},\
  }\href@noop {} {\bibfield  {journal} {\bibinfo  {journal} {Phys. Rev. Lett.}\
  }\textbf {\bibinfo {volume} {100}},\ \bibinfo {pages} {240601} (\bibinfo
  {year} {2008})}\BibitemShut {NoStop}%
\bibitem [{\citenamefont {Hoyos}\ and\ \citenamefont
  {Vojta}(2012)}]{HoyosVojta12}%
  \BibitemOpen
  \bibfield  {author} {\bibinfo {author} {\bibfnamefont {J.~A.}\ \bibnamefont
  {Hoyos}}\ and\ \bibinfo {author} {\bibfnamefont {T.}~\bibnamefont {Vojta}},\
  }\href {\doibase 10.1103/PhysRevB.85.174403} {\bibfield  {journal} {\bibinfo
  {journal} {Phys. Rev. B}\ }\textbf {\bibinfo {volume} {85}},\ \bibinfo
  {pages} {174403} (\bibinfo {year} {2012})}\BibitemShut {NoStop}%
\bibitem [{\citenamefont {Feynman}\ and\ \citenamefont
  {Hibbs}(1965)}]{FeynmanHibbs_book65}%
  \BibitemOpen
  \bibfield  {author} {\bibinfo {author} {\bibfnamefont {R.~P.}\ \bibnamefont
  {Feynman}}\ and\ \bibinfo {author} {\bibfnamefont {A.~R.}\ \bibnamefont
  {Hibbs}},\ }\href@noop {} {\emph {\bibinfo {title} {Quantum Mechanics and
  Path Integrals}}}\ (\bibinfo  {publisher} {McGraw-Hill},\ \bibinfo {address}
  {New York},\ \bibinfo {year} {1965})\BibitemShut {NoStop}%
\bibitem [{\citenamefont {Sachdev}(1999)}]{Sachdev_book99}%
  \BibitemOpen
  \bibfield  {author} {\bibinfo {author} {\bibfnamefont {S.}~\bibnamefont
  {Sachdev}},\ }\href@noop {} {\emph {\bibinfo {title} {Quantum phase
  transitions}}}\ (\bibinfo  {publisher} {Cambridge University Press},\
  \bibinfo {address} {Cambridge},\ \bibinfo {year} {1999})\BibitemShut
  {NoStop}%
\bibitem [{\citenamefont {Wolff}(1989)}]{Wolff89}%
  \BibitemOpen
  \bibfield  {author} {\bibinfo {author} {\bibfnamefont {U.}~\bibnamefont
  {Wolff}},\ }\href@noop {} {\bibfield  {journal} {\bibinfo  {journal} {Phys.
  Rev. Lett.}\ }\textbf {\bibinfo {volume} {62}},\ \bibinfo {pages} {361}
  (\bibinfo {year} {1989})}\BibitemShut {NoStop}%
\bibitem [{\citenamefont {Luijten}\ and\ \citenamefont
  {Bl{\"o}te}(1995)}]{LuijtenBlote95}%
  \BibitemOpen
  \bibfield  {author} {\bibinfo {author} {\bibfnamefont {E.}~\bibnamefont
  {Luijten}}\ and\ \bibinfo {author} {\bibfnamefont {H.~W.~J.}\ \bibnamefont
  {Bl{\"o}te}},\ }\href@noop {} {\bibfield  {journal} {\bibinfo  {journal}
  {Int. J. Mod. Phys. C}\ }\textbf {\bibinfo {volume} {6}},\ \bibinfo {pages}
  {359} (\bibinfo {year} {1995})}\BibitemShut {NoStop}%
\bibitem [{\citenamefont {Vojta}(2003{\natexlab{b}})}]{Vojta03b}%
  \BibitemOpen
  \bibfield  {author} {\bibinfo {author} {\bibfnamefont {T.}~\bibnamefont
  {Vojta}},\ }\href@noop {} {\bibfield  {journal} {\bibinfo  {journal} {J.
  Phys. A}\ }\textbf {\bibinfo {volume} {36}},\ \bibinfo {pages} {10921}
  (\bibinfo {year} {2003}{\natexlab{b}})}\BibitemShut {NoStop}%
\bibitem [{\citenamefont {Hrahsheh}\ \emph {et~al.}(2011)\citenamefont
  {Hrahsheh}, \citenamefont {Barghathi},\ and\ \citenamefont
  {Vojta}}]{HrahshehBarghathiVojta11}%
  \BibitemOpen
  \bibfield  {author} {\bibinfo {author} {\bibfnamefont {F.}~\bibnamefont
  {Hrahsheh}}, \bibinfo {author} {\bibfnamefont {H.}~\bibnamefont {Barghathi}},
  \ and\ \bibinfo {author} {\bibfnamefont {T.}~\bibnamefont {Vojta}},\
  }\href@noop {} {\bibfield  {journal} {\bibinfo  {journal} {Phys. Rev. B}\
  }\textbf {\bibinfo {volume} {84}},\ \bibinfo {pages} {184202} (\bibinfo
  {year} {2011})}\BibitemShut {NoStop}%
\bibitem [{\citenamefont {Young}\ and\ \citenamefont
  {Rieger}(1996)}]{YoungRieger96}%
  \BibitemOpen
  \bibfield  {author} {\bibinfo {author} {\bibfnamefont {A.~P.}\ \bibnamefont
  {Young}}\ and\ \bibinfo {author} {\bibfnamefont {H.}~\bibnamefont {Rieger}},\
  }\href@noop {} {\bibfield  {journal} {\bibinfo  {journal} {Phys. Rev. B}\
  }\textbf {\bibinfo {volume} {53}},\ \bibinfo {pages} {8486} (\bibinfo {year}
  {1996})}\BibitemShut {NoStop}%
\bibitem [{\citenamefont {Ubaid-Kassis}\ \emph {et~al.}(2010)\citenamefont
  {Ubaid-Kassis}, \citenamefont {Vojta},\ and\ \citenamefont
  {Schroeder}}]{UbaidKassisVojtaSchroeder10}%
  \BibitemOpen
  \bibfield  {author} {\bibinfo {author} {\bibfnamefont {S.}~\bibnamefont
  {Ubaid-Kassis}}, \bibinfo {author} {\bibfnamefont {T.}~\bibnamefont {Vojta}},
  \ and\ \bibinfo {author} {\bibfnamefont {A.}~\bibnamefont {Schroeder}},\
  }\href@noop {} {\bibfield  {journal} {\bibinfo  {journal} {Phys. Rev. Lett.}\
  }\textbf {\bibinfo {volume} {104}},\ \bibinfo {pages} {066402} (\bibinfo
  {year} {2010})}\BibitemShut {NoStop}%
\bibitem [{\citenamefont {Schroeder}\ \emph {et~al.}(2011)\citenamefont
  {Schroeder}, \citenamefont {Ubaid-Kassis},\ and\ \citenamefont
  {Vojta}}]{SchroederUbaidKassisVojta11}%
  \BibitemOpen
  \bibfield  {author} {\bibinfo {author} {\bibfnamefont {A.}~\bibnamefont
  {Schroeder}}, \bibinfo {author} {\bibfnamefont {S.}~\bibnamefont
  {Ubaid-Kassis}}, \ and\ \bibinfo {author} {\bibfnamefont {T.}~\bibnamefont
  {Vojta}},\ }\href@noop {} {\bibfield  {journal} {\bibinfo  {journal} {J.
  Phys. Condes. Matter}\ }\textbf {\bibinfo {volume} {23}},\ \bibinfo {pages}
  {094205} (\bibinfo {year} {2011})}\BibitemShut {NoStop}%
\bibitem [{\citenamefont {Demk\'o}\ \emph {et~al.}(2012)\citenamefont
  {Demk\'o}, \citenamefont {Bord\'acs}, \citenamefont {Vojta}, \citenamefont
  {Nozadze}, \citenamefont {Hrahsheh}, \citenamefont {Svoboda}, \citenamefont
  {D\'ora}, \citenamefont {Yamada}, \citenamefont {Kawasaki}, \citenamefont
  {Tokura},\ and\ \citenamefont {K\'ezsm\'arki}}]{Demkoetal12}%
  \BibitemOpen
  \bibfield  {author} {\bibinfo {author} {\bibfnamefont {L.}~\bibnamefont
  {Demk\'o}}, \bibinfo {author} {\bibfnamefont {S.}~\bibnamefont {Bord\'acs}},
  \bibinfo {author} {\bibfnamefont {T.}~\bibnamefont {Vojta}}, \bibinfo
  {author} {\bibfnamefont {D.}~\bibnamefont {Nozadze}}, \bibinfo {author}
  {\bibfnamefont {F.}~\bibnamefont {Hrahsheh}}, \bibinfo {author}
  {\bibfnamefont {C.}~\bibnamefont {Svoboda}}, \bibinfo {author} {\bibfnamefont
  {B.}~\bibnamefont {D\'ora}}, \bibinfo {author} {\bibfnamefont
  {H.}~\bibnamefont {Yamada}}, \bibinfo {author} {\bibfnamefont
  {M.}~\bibnamefont {Kawasaki}}, \bibinfo {author} {\bibfnamefont
  {Y.}~\bibnamefont {Tokura}}, \ and\ \bibinfo {author} {\bibfnamefont
  {I.}~\bibnamefont {K\'ezsm\'arki}},\ }\href {\doibase
  10.1103/PhysRevLett.108.185701} {\bibfield  {journal} {\bibinfo  {journal}
  {Phys. Rev. Lett.}\ }\textbf {\bibinfo {volume} {108}},\ \bibinfo {pages}
  {185701} (\bibinfo {year} {2012})}\BibitemShut {NoStop}%
\end{thebibliography}%

\end{document}